\documentclass{amsart}

\usepackage{graphicx}
\usepackage{color}
\newtheorem{theorem}{Theorem}[section]
 \newtheorem{lemma}{Lemma}[section]
 \newtheorem{prop}{Proposition}[section]

  \newtheorem{rem}{Remark}[section]

\UseRawInputEncoding
\usepackage[russian, english]{babel}

\def\dfrac#1#2{\displaystyle{#1\over #2}}
\def\za{\alpha}
\def\zb{\beta}
\def\zp{\varphi}

\def\zt{\theta}
\def\bv{{\bf v}}
\def\bp{{\bf p}}

\def\Div{\mbox{div}\,}

\def\bB{{\bf B}}

\def\bE{{\bf E}}

\begin{document}

\markboth{Rozanova, Chizhonkov}{Stabilization and blow-up in 
a collisional plasma}

\title[Stabilization and blow-up in a
collisional plasma]{Stabilization and blow-up in the  relativistic model of  cold
collisional plasma}

\author[Rozanova]{Olga S. Rozanova}
\address{%
Olga S. Rozanova\\
Mathematics and Mechanics Department\\ Lomonosov Moscow State University\\ Leninskie Gory
Moscow 119991\\
Russian Federation}
\email{rozanova@mech.math.msu.su}

\author[Chizhonkov]{Eugeniy V. Chizhonkov}
\address{ %
Eugeniy V. Chizhonkov\\
Mathematics and Mechanics Department\\ Lomonosov Moscow State University\\ Leninskie Gory
Moscow 119991\\
Russian Federation}
\email{chizhonk@hotmail.com}
\thanks{Partially supported by the Moscow Center for Fundamental and Applied Mathematics.}

%
%


\begin{abstract}
We study the influence of the factor of electron-ion collisions on the solution of the Cauchy problem in the one-dimensional relativistic model of cold plasma and show that, depending on their intensity and initial data, two scenarios are possible: either the solution remains smooth and stabilizes to a stationary state, or during a finite time the oscillations blowup.
 In contrast to the nonrelativistic model, when exact conditions can be obtained separating the two behaviors,
in a much more complicated relativistic situation, it turns out to be possible to analytically estimate from below the time during which the existence of a smooth solution and the guaranteed number of oscillations during this time. In addition, we show that in contrast to the relativistic case without taking into account collisions, when oscillations corresponding to arbitrarily small deviations from the zero equilibrium position blow up, the presence of electron collisions can suppress the blow-up of sufficiently small oscillations. Further, based on the analysis of characteristics, a numerical algorithm is constructed, the order of accuracy of which is determined only by the smoothness of the initial data. Numerical experiments are presented to illustrate the theoretical results. The initial conditions are chosen as reasonably as possible from the point of view of full-scale physical experiments.
\end{abstract}





\maketitle
\section
{ Introduction}\label{S1}

The hydrodynamic model of cold plasma is one of the most commonly used and has long been used in physics \cite{GR75}, \cite{david72}.
Depending on what effects we want to study, the influence of ion motion can be neglected or taken into account to one degree or another.
One of the specific features of the cold plasma model is that it has physical meaning only for smooth solutions. It is believed that when the singularities of the solution are formed (here these are delta-shaped singularities in the density component), energy is released, which leads to the "heating" of the plasma and the need to replace the model.
Indeed, the formation of a singularity within the framework of the Lagrangian approach corresponds to the intersection of adjacent trajectories.
When two different particles occupy the same position in space and time, then further tracking their motion requires the use of more complex models than classical electrodynamics, since the infinite concentration of electric charge requires a special interpretation.

For the nonrelativistic approximation, the problem of the criterion for the formation of a singularity in the solution of the collisional cold plasma equations in terms of the initial data is completely solved in \cite {RChD20} (the same problem in terms of the Euler-Poisson equations is considered in \cite {Tadmor}).
It turned out that for a constant collision coefficient, one can strictly divide the initial data into those that in a finite time lead to the formation of a singularity and those that correspond to a solution that is globally smooth in time.
In the latter case, the solution asymptotically tends to a stationary state as $ t \to \infty $.
In addition, there is a threshold value of the collision coefficient $ \nu> 0 $, after which the oscillatory nature of the damping is replaced by a monotonic one (this value is $ \nu = 2 $, while the physically natural values of $ \nu $ are much smaller).
However, for an arbitrarily large fixed value of the collision coefficient, there are initial data at which a singularity is formed.
 From the point of view of physical applications, results of this kind are of limited value, since nonrelativistic models are regarded as a strong simplification.

It is believed that relativistic plasma oscillations are always blow up in the general case. As shown in \cite {RChZAMP21}, for the collisionless case, this is to a certain extent true, since nonlinear resonance occurs in the solution (except for solutions that correspond to special initial data). In particular, even an arbitrarily small perturbation of the trivial equilibrium position generally leads to blowing up.

A natural question arises: will the presence of collisions be able to extend the lifetime of the solution, or, perhaps, completely suppress the blow-up? In the present work, we obtain a positive answer to this question, extending the results of  \cite{RChZAMP21}. We show that the lifetime of a smooth solution can be estimated from below, including in terms of the number of oscillations, and this time increases as $ \nu $ grows. Moreover, for any fixed initial data, choosing a sufficiently large $ \nu> 0 $ allows one to obtain a global in time smooth solution stabilizing to a trivial constant state. Further, we will show that for each arbitrarily small $ \nu> 0 $, there exists such a small perturbation of the stationary state that the corresponding solution remains globally smooth.

It should be noted that the influence of collisions on plasma oscillations has already been analyzed by other authors (see, for example,  \cite{infeld},\cite{verma}). However, in these works, the plasma resistance was taken into account simultaneously with the viscosity. As a result, the blow-up effect fell outside the attention of researchers. For the first time, it was possible to trace the influence of electron collisions on the breaking of plane plasma oscillations in  \cite{FrCh}. However, there the analysis was carried out by asymptotic and numerical methods, therefore, it was not possible to obtain exact mathematical formulations about the conditions of blowup (or, conversely, non-blowup).

The article is structured as follows. Section \ref{S2} contains a detailed statement of the problem in Eulerian variables and a reduction to a quasilinear system of hyperbolic equations. Section \ref{S3} contains the analytical results announced above.
Further, the method for proving analytical statements is transformed into a numerical algorithm, the order of accuracy of which is determined only by the smoothness of the solution. The paper also contains numerical experiments that illustrate the theoretical results. The initial conditions are chosen as reasonable as possible from the point of view of full-scale physical experiments. For the sake of completeness, the physical substantiation of the intensity of electron collisions is given. In the conclusion, the results of the research are systematized.

The main goal of this work is to substantiate and demonstrate the fact that an increase in the parameter $ \nu $ leads to an expansion of the range of initial data for which there is a smooth global solution in time.
\section
{ Formulation of the problem}\label{S2}
We will consider plasma as a compressible relativistic electron liquid, neglecting recombination effects and ion motion. In vector form, the system of hydrodynamic equations describing it, together with Maxwell's equations, is
\begin{equation}
\label{base1}
\begin{array}{l}
\dfrac{\partial n }{\partial t} + \Div(n \bv)=0\,,\quad
\dfrac{\partial \bp }{\partial t} + \left( \bv \cdot \nabla \right) \bp
= e \, \left( \bE + \dfrac{1}{c} \left[\bv \times  \bB\right]\right) - \nu_{ei} \bp,\vspace{0.5em}\\
\gamma = \sqrt{ 1+ \dfrac{|\bp|^2}{m^2c^2} }\,,\quad
\bv = \dfrac{\bp}{m \gamma}\,,\vspace{0.5em}\\
\dfrac1{c} \frac{\partial \bE }{\partial t} = - \dfrac{4 \pi}{c} e n \bv
 + {\rm rot}\, \bB\,,\quad
\dfrac1{c} \frac{\partial \bB }{\partial t}  =
 - {\rm rot}\, \bE\,, \quad \Div \bB=0\,,
\end{array}
\end{equation}
where
$e, m$ are the charge and mass of an electron (here the electron charge has a negative sign:$e < 0$),
$ c $ is the speed of light;
$ n, \bp, \bv$ are the density, momentum and speed of electrons;
$\gamma$ is the Lorentz factor;
$ \bE, \bB $ are vectors of electric and magnetic fields.

The system of equations (\ref {base1}) is one of the simplest models of plasma, which is often called the equations of hydrodynamics of "cold" plasma; it is well known and described in sufficient detail in textbooks and monographs (see, for example,~\cite {S71} - \cite {SR12}).

 The equation for the momentum contains the term $ \nu_{ei} \bp $, which describes electron-ion collisions.
Taking this effect into account can be interpreted as the force of friction between particles; in the non-relativistic case (see, for example,~\cite {ABR78})
they often use a formula like
$$
- \nu_{\za\zb} \left( \bv_{\za} - \bv_{\zb} \right),
$$
where $ \nu_{\za \zb} $ is the effective frequency of collisions of charged particles of the type $ \za $ with particles of the type $ \zb $ when $ \za \neq \zb $.
For stationary ions ($ \bv_{\zb} = 0 $) the formula is simplified.

In order to analyze and construct a numerical solution of plane one-dimensional relativistic plasma oscillations taking into account collisions
the basic equations~(\ref{base1}) can be greatly simplified.

We will denote the independent variables in the Cartesian coordinate system
in the usual way --- $ (x, y, z) $, and assume that
\begin{itemize}
\item the solution is determined only
$ x- $ components of vector functions ${\bp}, {\bv}, {\bE};$
\item there is no dependence in these functions on the variables $ y $ and $ z $.
\end{itemize}

Then from the system \eqref {base1} we get
\begin{equation}
\begin{array}{c}
\dfrac{\partial n }{\partial t} +
\dfrac{\partial }{\partial x}
\left(n\, v_x \right)
=0,\quad
\dfrac{\partial p_x }{\partial t} + v_x \dfrac{\partial p_x}{\partial x}= e \,E_x -\nu_{ei} \,p_x,
\vspace{1 ex}\\
\gamma = \sqrt{ 1+ \dfrac{p_x^2}{m^2c^2}}\,, \quad
{v_x} = \dfrac{p_x}{m \,\gamma}, \quad
\dfrac{\partial E_x }{\partial t} = - 4 \,\pi \,e \,n\, v_x\,.
\end{array}
\label{3gl2}
\end{equation}

We introduce the dimensionless quantities
$$
\rho = k_p x, \quad \theta = \omega_p t, \quad
V = \dfrac{v_x}{c}, \quad
P = \dfrac{p_x}{m\,c}, \quad
E = -\,\dfrac{e\,E_x}{m\,c\,\omega_p}, \quad
N = \dfrac{n}{n_0}, \quad \nu = \dfrac{\nu_{ei}}{\omega_p},
$$
 where $\omega_p = \left(4 \pi e^2n_0/m\right)^{1/2}$  is the plasma frequency, $n_0$ is the value of the unperturbed electron density, $k_p = \omega_p /c$.
In the new variables \eqref{3gl2} takes the form
\begin{equation}
\begin{array}{c}
\dfrac{\partial N }{\partial \theta} +
\dfrac{\partial }{\partial \rho}
\left(N\, V \right)
=0,\quad
\dfrac{\partial P }{\partial \theta} + E +
V \dfrac{\partial P}{\partial \rho} + \nu P= 0, \vspace{1.5 ex}\\
\gamma = \sqrt{ 1+ P^2}, \quad
V = \dfrac{P}{\gamma},\quad
\dfrac{\partial E }{\partial \theta} = N\, V\,.
\end{array}
\label{3gl3}
\end{equation}

From the first and last equations \eqref {3gl3} it follows
$$
\dfrac{\partial }{\partial \theta}
\left[ N +
\dfrac{\partial }{\partial \rho} E \right] = 0.
$$
This relationship is valid both in the absence of plasma oscillations and in their presence. Therefore, under the traditional assumption of a uniform background charge density of stationary ions, this implies a simpler expression for the electron density $N(\rho,\theta)$:
\begin{equation}
 N(\rho,\theta) = 1 -
\dfrac{\partial  E(\rho,\theta) }{\partial \rho}.
\label{3gl4}
\end{equation}
Formula \eqref {3gl4} is a special case of the Gauss theorem~\cite {david72} pp.33-53, which in differential dimensional form has the form
$ \Div \bE =  4\,\pi\,e (n - n_0).$
Using \eqref {3gl4} in \eqref {3gl3}, we arrive at the equations describing plane one-dimensional relativistic plasma oscillations taking into account collisions:
\begin{equation}\label{u1}
\dfrac{\partial P }{\partial \theta}+
V\,\dfrac{\partial P}{\partial \rho}  + E + \nu P = 0,\quad
\dfrac{\partial E }{\partial \theta}  +
V\,\dfrac{\partial E}{\partial \rho}- V = 0,
\quad V = \dfrac{P}{\sqrt{1+P^2}}\,.
\end{equation}
Here $ \rho $ and $ \theta $ are dimensionless coordinates with respect to
space and time, respectively. The variable $ P $ describes the momentum of the electrons, $ V $ is the velocity of the electrons, and $ E $ is a function that characterizes the electric field.
To the system \eqref{u1} it is necessary to add the equation \eqref {3gl4}, which describes the behavior of  the electron density, the most important function in the cold plasma model.
Below we  study in the half-plane $\{(\rho,\theta)\,:\, \rho \in {\mathbb R},\; \theta
> 0\}$ a solution of the Cauchy problem for \eqref{3gl4}, \eqref{u1} with initial conditions
\begin{equation}\label{cd1}
     P(\rho,0) = P_0(\rho), \quad
     E(\rho,0) = E_0(\rho), \quad
 \rho \in {\mathbb R}.
\end{equation}

System \eqref {u1} is of the hyperbolic type.
It is well known that for such systems there exists, locally in time, a unique solution to the Cauchy problem of the same class as the initial data~\cite {Daf16}. Below, it is enough for us to require the smoothness of the initial data $ C^2 (\mathbb R) $.
Since the coefficient of electron collisions in real problems is small, we  mainly consider $ \nu \in [0,2) $ (in \cite {RChD20} a detailed explanation of how the upper threshold value of $ \nu $ arises).

 System \eqref{u1} for functions $P(\rho,\theta)$ è $E(\rho,\theta)$ can be considered together with an extended system for derivatives
$$
Q(\rho,\theta) = \dfrac{\partial P(\rho,\theta)}{\partial \rho}, \quad D(\rho,\theta) = \dfrac{\partial E(\rho,\theta)}{\partial \rho},
$$
namely,
\begin{equation}\label{du1}
\dfrac{\partial Q }{\partial \theta} + V\,\dfrac{\partial Q}{\partial \rho} +D + \dfrac{Q^2}{(1+P^2)^{3/2}} + \nu Q =0, \quad
\dfrac{\partial D }{\partial \theta} + V\,\dfrac{\partial D}{\partial \rho} = (1-D)\dfrac{Q}{(1+P^2)^{3/2}},
\end{equation}
with initial conditions
\begin{equation}\label{dcd1}
     Q(\rho,0) = \dfrac{d\, P_0(\rho)}{d\, \rho}, \quad D(\rho,0) = \dfrac{d\, E_0(\rho)}{d\, \rho}, \quad
		 \rho \in {\mathbb R}.
\end{equation}
The extended system is useful for studying the formation of singularities of the solution associated with infinite derivatives.

We consider the characteristics of the system~\eqref{u1}, outgoing from a fixed point $\rho_0$. The system of characteristics has the form
\begin{equation}\label{char}
     \dfrac{dP}{d\theta}=-E - \nu\,P,\quad \dfrac{dE}{d\theta}=\dfrac{P}{\sqrt{1+P^2}},\quad \dfrac{d\rho}{d\theta}=\dfrac{P}{\sqrt{1+P^2}},
\end{equation}
whence it immediately follows that
\begin{equation}\label{first_int}
2\sqrt{1+P^2}+E^2 \le
2\sqrt{1+P^2_0(\rho_0)}+E^2_0(\rho_0)= {\mathcal E}_0(\rho_0)=\mbox{\rm const}.
\end{equation}
Thus, the solution itself is always bounded, which cannot be guaranteed for its derivatives.

 Derivatives along a characteristic starting from a point $\rho_0$, satisfy
\begin{equation}\label{char1d}
     \dfrac{dQ}{d\theta}=-\nu Q - D-K Q^2,\quad \dfrac{dD}{d\theta}=K(1-D){Q},\quad K=(1+P^2)^{-3/2}.
     \end{equation}
In the nonrelativistic case, system \eqref{char1d}  is split off from \eqref{char}, but in the relativistic case they are connected via
$ K (\theta) $, that is, $ P (\theta) $. This complicates the problem considerably.

 Although we do not know the explicit representation of $ K (\theta) $, we can evaluate this function from two sides. Namely,
 \begin{equation}\label{estK}
0 < \dfrac{8}{{\mathcal E}_0^3} = K_-\le K(\theta)\le K_+ =1.
\end{equation}
The upper bound for $ K_+ $ is obvious,  the lower bound follows from \eqref {first_int}.

Let us note that the system \eqref{u1} possesses solutions in the form of simple waves, i.e. with the constraint $E=E(P)$. For this particular case, system \eqref{u1} reduces to one quasilinear equation, which can be analyzed separately. However, this analysis is not so simple as for $\nu=0$, when one can obtain an explicit criterion of singularities formation (see \cite{RChZAMP21}). The simple waves for $\nu>0$ satisfy the equation
\begin{equation}\label{SW}
  \frac{dE(P)}{dP}=-\frac{P}{\sqrt{1+P^2} (E(P)+\nu P)},
\end{equation}
which cannot be integrated explicitly.

\section
{  Analytical results}\label{S3}

The results on estimates of the lifetime of a smooth solution to the equations of relativistic cold plasma, proved in this section, continue the method developed  in \cite {RChZAMP21} (see also \cite {RChDAN20}).

Let us summarize it briefly. Since the formation of a singularity is associated with infinite derivatives of the solution of \eqref {u1}, the solution remains smooth as long as the projection of the characteristic curve described by  equations \eqref {char}, \eqref {char1d} onto the phase plane $ (D , Q) $ remains bounded. We denote by $ \za = D (\rho, 0), \zb = Q (\rho, 0) $ the starting point of this projection. As shown in \cite {RChZAMP21}, in the nonrelativistic case, when system \eqref {char1d}  is autonomous, for $ \nu = 0 $
condition
\begin{equation}\label{cond0}
 \zb^2 + 2\, \za-1< 0
\end{equation}
means that the phase curve in the $ (D, Q) $ plane is an ellipse, which ensures that the derivatives are bounded in time.
If the condition \eqref {cond0} is not met, the curve on the phase plane is a parabola or hyperbola, and its unboundedness corresponds to the fact that the derivatives become infinite in a finite time. In this case, we are dealing with a criterion for the formation of a singularity in terms of the initial data. For $ \nu> 0 $, in the nonrelativistic case one can also obtain a criterion of this kind \cite {RChD20}, which is rather cumbersome. However, its analysis shows that the condition \eqref {cond0} enough to keep the phase trajectory bounded. For $ \nu> 0 $, the equilibrium position $ (0,0) $ is asymptotically stable and, globally in time, the smooth solution $ (V, E) $ tends to
trivial stationary state at $ \theta \to \infty $.

In this case, the behavior of the solution for $ \nu \in [0,2) $ and $ \nu> 2 $ is sharply different. At $ \nu \in [0,2) $, the equilibrium position $ (0,0) $ is the focus, therefore the phase trajectory makes an infinite number of revolutions around the origin, which corresponds to the oscillatory motion of the medium. For $ \nu> 2 $, the equilibrium position $ (0,0) $ is a node, and the medium does not oscillate. In this case, the phase trajectory always turns out to be in the quadrant $ D <0 $, $ Q <0 $, it either goes to the origin of coordinates in infinite time, or goes to infinity in a finite time. The latter corresponds to the formation of a singularity of the solution, but by increasing $ \nu $, any fixed initial data can be translated into an area such that the trajectory leaving it goes to the origin. Since the physically natural value of $ \nu $ is small, no suppression of oscillations is observed in experiments. However, formally, for arbitrary initial data \eqref {cd1}, one can find such $ \nu $ that the solution of the problem \eqref {u1}, \eqref {cd1} will be globally smooth in time.

In the relativistic case, such results are impossible due to the fact that the system \eqref {char1d} ceases to be autonomous. Therefore, one has to be content with estimates of the location of the projections of the phase curves on $ (Q, D) $, based on the fact that the  solution of \eqref {char1d} can be estimated through its analogue with the substitution $ K_\pm = \rm const $ instead of $ K = K ( \theta) $. Note that $ K (\theta) $ can be found in quadratures from \eqref {char}, but this does not help the investigation, since $ K (\theta) $ cannot be explicitly represented. Immediately, we note that the smaller the difference $ K_+ - K_- = 1-K_- $, that is, the less the initial data differ from the zero equilibrium position, the more accurate estimates can be obtained. For the nonrelativistic case, this difference is zero.

Let us assume that the solution do not correspond to a simple wave given as \eqref{SW} and introduce new variables $u={D}/{Q}$, $\lambda=(1-D)/{Q}$, in which the system \eqref {char1d} is written as
\begin{equation}\label{supp1}
 \dfrac{du}{d\theta}=u^2+\nu u+K,\quad \dfrac{d\lambda}{d\theta}=\lambda u+\nu \lambda.
\end{equation}

 From the first equation \eqref {supp1} it follows that $ u $ can tend to infinity in a finite time (or should do so depending on the choice of $ \nu $). This can happen for two reasons: $ Q $ vanishes for finite $ D $ or $ Q $ and $ D $ both go to infinity. The latter situation can also occur if $ u $ is bounded. Similarly, if $ u $ vanishes, then $ Q $ tends to infinity or $ D $ vanishes. Analysis of the phase trajectories shows that for $ D \ge 1 $, the value of $ Q $ always tends to infinity in a finite time, but from the requirement that the density be positive it follows that $ D <1 $. In the domain $ D \in (0,1), \, Q> 0 $, the trajectory reaches the boundary $ Q = 0 $, and in the domain $ D \in (0,1), \, Q <0 $ and $ D <0, \, Q> 0 $ the trajectory reaches the boundary $ D = 0 $. In the quadrant $ D <0, \, Q <0 $, where $ u> 0 $, the value of $ u $ can go to infinity for various reasons. If $ Q $ turns to zero, then the trajectory makes a revolution around the origin, otherwise $ Q $ turns to infinity in a finite time. We need  to distinguish  these two situations.

Let us introduce the following notation:
 $$ \mbox{quadrant \, I}:  D <0, \, Q >0,\quad \mbox{quadrant\, II}:  D >0, \, Q >0,$$$$    \mbox{quadrant \, III}:  D >0, \, Q <0,\quad
 \mbox{quadrant\, IV}:  D <0, \, Q <0.
  $$

It follows from the above that the derivatives can become unbounded only in   quadrant IV. We will say that a trajectory makes a revolution (the solution makes one oscillation) if it returns to the same quadrant from which it left.

For a simple wave, $ u = E'(P) $ (see \eqref{SW}) and the behavior of $ u $ does not correspond to the behavior of the derivatives. This case requires a separate analysis.

\subsection{Behavior on the first revolution}

First, we obtain the simplest sufficient conditions for the initial data, allowing us to conclude that the trajectory certainly makes one revolution (returns from quadrant IV  to  quadrant I) or goes to infinity in quadrant  IV.

The behavior of the solution for $ \nu \in [0,2) $ and $ \nu\ge 2 $ is sharply different. In the first case, we encounter nonlinear resonance, so the phase trajectory can go to infinity both at the first and at subsequent revolutions. In the second case, there is no oscillations and the behavior of the solution is determined already on the first revolution.

\begin{prop}\label{utv2.1}
Let $ 0 <\nu <2 $, and the initial data \eqref {cd1}  are bounded on $ \mathbb R $, belong to the class $ C^2 (\mathbb R) $ and  do not correspond to a simple wave given as \eqref{SW}.
  If condition \eqref {cond0} holds for all $ \rho $
together with any of the conditions
\begin{itemize}
\item $ \zb <0 $
or
\item $ \zb = 0 $, $ \za> 0 $,
\end{itemize}
then
  the solution to  problem \eqref {u1}, \eqref {cd1} is classically smooth at least until the time $ t_* = \min \limits_{\rho \in \mathbb R} T (\rho) $, $ t_* \ge 2 \pi $ (each trajectory makes at least one revolution).
 \end{prop}


Let us denote
\begin{equation}\label{Tminus}
 T_-=\frac{1}{K_--\nu^2/4} \left(\frac{\pi}{2} -  \arctg \frac{\za/\zb +\nu/2}{\sqrt{K_--\nu^2/4}}    \right).
\end{equation}


\begin{prop}\label{utv2.2}
Let $ 0 <\nu <2 \sqrt {K_-} $ (see (\ref {estK})), the initial data (\ref {cd1})  are bounded on $ \mathbb R $, belong to the class $ C^2 (\mathbb R) $ and  do not correspond to a simple wave given as (\ref{SW}).
Then if for some $ \rho_0 $ condition
\begin{equation}\label{cond_bl}
\beta^2+\frac{\exp\{\nu T_-\}}{ K_-}(2 \alpha-1)+\left(\frac{1}{K_{-}-\nu^2/4}-\frac{\exp\{\nu T_-\}}{ K_-}  \right)\alpha^2>0,
 \end{equation}
 holds, then the solution to problem (\ref{u1}), (\ref{cd1})  loses its smoothness before $t_*=\min\limits_{\rho\in \mathbb R} T(\rho)$, $t_* \ge 2\pi$
(there is a trajectory that does not make a revolution).
\end{prop}

\proof We prove Propositions \ref{utv2.1} and \ref{utv2.2} simultaneously. First of all, note that  ${Q}^{-1} = \lambda+u $ and this expression for $ Q <0 $ can be estimated from both sides as
\begin{equation}\label{theta_pm}
 \psi_-(\theta)\le \frac{1}{Q}\le \psi_+(\theta),
\end{equation}
where
$$
 \psi_{\pm}(\theta)=-\frac{\nu}{2}+{\tilde K}_{\pm}\tg \zp_{\pm}
+ \exp\{\nu\theta/2\}
\dfrac{\lambda_0\sqrt{1+\tg^2 \zp_{\pm}}}{\sqrt{\dfrac{(u_0+\nu/2)^2}{{\tilde K}_{\pm}^2}+1}},
$$
$$
{\tilde K}_{\pm} = \sqrt{K_{\pm}-\nu^2/4},
\quad \zp_{\pm} = {\tilde K}_{\pm}\,\theta+\arctg \frac{u_0+\nu/2}{{\tilde K}_{\pm}}.
$$
If in the domain of negative $ Q $, that is, in   quadrants III and IV, $ {Q}^{- 1} $ vanishes before it turns to infinity, then the solution loses its smoothness in a finite time. Thus, we need to find the condition for $ \psi_- (\theta) $ to vanish when  $ \psi_- (\theta) $ increases with respect to  $\theta$.

On the other hand, if
  $ \psi _ + (\theta)\ne 0 $ for $\theta\in(0,T_+)$, where
 $$
T_+=\dfrac{1}{1-\nu^2/4} \left(\frac{\pi}{2} -  \arctg \dfrac{u_0+\nu/2}{{\tilde K}_{+}}    \right),
$$
  then the trajectory  returns to quadrant I and the solution remains smooth for at least one revolution of the trajectory around the origin.
  Direct calculation shows that
 $$
 \psi_+(\theta)= \frac{\exp\{\nu\theta/2\}}{\sqrt{\dfrac{(u_0+\nu/2)^2}{{\tilde K}_{+}^2}+1}}\,F_+(\theta) ,
$$
 where $F_+(\theta)$ characterizes the denominator $Q$ (see \cite{RChD20}) and is defined as
$$
  F_+(\zt)= 1-\za+ \left(\dfrac{\nu \za+2 \zb}{2 {\tilde K}_{+} } \,\sin {\tilde K}_{+} \zt + \za \,\cos {\tilde K}_{+} \zt \right)\, \exp(-\nu \zt/2).
 $$
Thus, the conditions guaranteeing the smoothness of the solution during the first oscillation correspond to the conditions of non-vanishing of  $F_+(\theta)$.
It follows from the results of \cite {RChD20} that non-vanishing of $ F_+ (\theta) $ for $ \nu = 0 $ guarantees non-vanishing of this function for an arbitrary
$ 0 <\nu <2 $, therefore the inequality $ \zb^2 + 2 \, \za-1 <0 $ leads to the desired result. Conditions $ \zb <0 $ or $ \zb = 0 $, $ \za> 0 $ ensures that the trajectory falls into quadrant III. Note that if this condition is not imposed, then in the absence of an upper bound for the trajectory for $ Q> 0 $ (in quadrants I and II), it cannot be guaranteed that the condition \eqref {cond0} will be satisfied when the trajectory is in quadrant III. If the trajectory starts, for example, in  quadrant I, then it can go to infinity in quadrant IV and not make a full revolution. Note that the results of the next section allow one to obtain two-sided estimates of the position of the trajectory.
In conclusion, note that in \cite {RChZAMP21} it is shown that a turn along a trajectory for $ K \ge 1 $ takes time greater than or equal to $ 2 \pi $.
Proposition \ref{utv2.1} is proved. $\Box$
\medskip

 Let us proceed to the proof of Proposition \ref{utv2.2}. We will restrict ourselves here to a rather rough but simple construction. More accurate construction of a domain on the plane corresponding to the initial data leading to the loss of smoothness  is described below, in Remark \ref{rem2.1}.
   Here we note that since $ \lambda_0 <0 $, then
 $$
\psi_-(\theta)>\bar\psi_-(\theta)=-\frac{\nu}{2}+{\tilde K}_{-}\tg \zp_-
+ \exp\{\nu T_{-}/2\}
\dfrac{\lambda_0\sqrt{1+\tg^2 \zp_-}}{\sqrt{\dfrac{(u_0+\nu/2)^2}{{\tilde K}_{-}^2}+1}},
$$
where $T_-$ is defined by \eqref{Tminus}, and $\za/\zb = u_0$.

It is easy to calculate that $\bar\psi_-(\theta)$ has a root on $(0,T_-)$, if
$$
 K_--\exp\{\nu T_-\}\,
\frac{{\tilde K}_{-}^2}{(u_0+\nu/2)^2+{\tilde K}_{-}^2}\,\lambda^2_0>0,
$$
which can be rewritten as \eqref{cond_bl}.
Proposition \ref{utv2.2} is proved. $\Box$

\bigskip

 \begin{rem}\label{rem2.1}
 One can see that
  $$
  \psi_-(\theta)= \frac{\exp\{\nu\theta/2\}}{\sqrt{\dfrac{(u_0+\nu/2)^2}{{\tilde K}_{-}^2}+1}}\,F_-(\theta) , $$
$$	
F_-(\theta)=1-\alpha +\left( \frac{\nu \alpha+2 \beta}{{\tilde K}_{-}} \sin ({\tilde K}_{-}\theta)+\alpha \cos({\tilde K}_{-}\theta) \right)
\exp(-\nu \zt/2),
$$
therefore it is possible to construct on the plane $ (\alpha, \beta )$ the curve $ \Phi_\nu (\alpha, \beta) = 0 $
 (the procedure is described in detail in \cite{RChD20}). The initial data corresponding to a solution that loses its smoothness already during the first revolution of the phase trajectory  are to the right of this curve. Since we are dealing only with an estimate of $ Q $, we cannot divide the plane $ (\alpha, \beta) $ into parts corresponding to a smooth and losing smoothness solution exactly (to obtain a criterion for the formation of a singularity).
\end{rem}

For the case $ \nu> 2 $, the estimate \eqref {theta_pm} is also valid, but here
$$
 \psi_{\pm}(\theta)=-\frac{\nu}{2}+{\tilde K}_{\pm}\tanh \zp_{\pm}
+ \exp\{\nu\theta/2\}
\dfrac{\lambda_0\sqrt{1+\tanh^2 \zp_{\pm}}}{\sqrt{\dfrac{(u_0+\nu/2)^2}{{\tilde K}_{\pm}^2}+1}},
$$
$$
{\tilde K}_{\pm} = \sqrt{\nu^2/4-K_{\pm}},
\quad \zp_{\pm} = {\tilde K}_{\pm}\,\theta+{\rm arcth} \frac{u_0+\nu/2}{{\tilde K}_{\pm}}.
$$

Based on these estimates, one can obtain statements similar to the previous two, but here a statement of a different kind is also true.

\begin{prop}\label{utv2.3}
For any initial data \eqref {cd1} (not corresponding to a simple wave given as \eqref{SW})  belonging to the class $ C^2 (\mathbb R) $ and bounded at $ \mathbb R $
there exists a value $ \nu> 0 $ such that the solution to  problem \eqref {u1}, \eqref {cd1}  remains classically smooth for all $ t> 0 $ and stabilizes to a stationary state for $ t \to \infty $.
 \end{prop}

\proof

Arguing as in the proof of Proposition \ref{utv2.1}, we obtain that for $\nu>2$
 $$
 \psi_+(\theta)= \frac{\exp\{\nu\theta/2\}}{\sqrt{\dfrac{(u_0+\nu/2)^2}{{\tilde K}_{+}^2}+1}}\,F_+(\theta) ,
$$
 $$
  F_+(\zt)= 1-\za+ \left(\dfrac{\nu \za+2 \zb}{2 {\tilde K}_{+} } \,\sinh {\tilde K}_{+} \zt + \za \,\cosh {\tilde K}_{+} \zt \right)\, \exp(-\nu \zt/2)
 $$
 (see \cite{RChD20}), and the conditions guaranteeing the smoothness of  solution correspond to the conditions of non-vanishing of $F_+(\theta)$. It is easy to see that for any fixed $ \alpha $ and $ \beta $ by increasing $ \nu $ one can get non-vanishing of  $ F_ + (\theta) $ for all $ t> 0 $.  Moreover, $ D $ and $ Q $ tend to zero as $ t \to \infty $. Proposition \ref{utv2.3} is proved. $\Box$

\bigskip

\subsection{Calculation of the guaranteed number of revolutions}

In this section, we describe a procedure that allows us to obtain an estimate from below for the number of revolutions that makes projections onto the plane $ (D, Q) $ of the trajectory going from a particular point $ \rho_0 $ before the formation of a singularity, and thus an estimate of the time of formation  of a singularity from below as infimum over $ \rho_0 \in \mathbb {R} $. The procedure consists in constructing the constraints of the projection of the trajectory, that is, the curves between which this projection is guaranteed to be located. Especially, it is important to construct the upper limiter, since if the condition \eqref {cond0} is not satisfied for it, then this condition cannot be guaranteed for the projection of the trajectory itself. The equations of the curves constituting the limiter are found explicitly. As mentioned above, this procedure makes sense for small $ \nu $.

Note that \eqref {char1d} implies the non-autonomous equation
\begin{equation}\label{QDP}
  \frac12   \dfrac{d Q^2}{d D}=-\nu \frac{Q}{(1-D) K(P)} - \frac{D}{(1-D) K(P)}-\frac{Q^2}{(1-D) }:=\Psi(D,Q,P).
     \end{equation}
    Since the density is positive, it follows that $ 1-D> 0 $. Let us denote
      \begin{eqnarray*}
       \Sigma_{ij} &=& -\nu \frac{Q}{(1-D) K_i} - \frac{D}{(1-D) K_j}-\frac{Q^2}{(1-D)},
     \end{eqnarray*}
 where $ i $, $ j $ take the values $ m $ or $ p $, and $ K_m = K _- $, $ K_p = K _ + $.
      Thus,
     \begin{eqnarray*}
   \Sigma_{pm}(D,Q)\le   \Psi(D,Q,P)\le \Sigma_{mp}(D,Q), &\quad (Q>0, D<0), \, I \, \mbox{êâàäðàíò}, \\   \Sigma_{pp}(D,Q)\le   \Psi(D,Q,P)\le \Sigma_{mm}(D,Q),& \quad (Q>0, D>0), \, II \, \mbox{êâàäðàíò},\\
    \Sigma_{mp}(D,Q)\le   \Psi(D,Q,P)\le \Sigma_{pm}(D,Q), &\quad (Q<0, D>0), \, III \, \mbox{êâàäðàíò}, \\   \Sigma_{mm}(D,Q)\le   \Psi(D,Q,P)\le \Sigma_{pp}(D,Q),& \quad (Q<0, D<0), \, IV \, \mbox{êâàäðàíò}.
     \end{eqnarray*}
In what follows, we denote the integral curves of the equations
$\dfrac{d Q^2}{d D}=2\Sigma_{ij}(D,Q) $,   (i.e. \eqref{QDP}) as $\Lambda_{ij}$, where $ij$ takes indices $mm$, $mp$, $pm$, $pp$.
Since in
quadrants I and II the function $ D $ increases as it moves along the projection of the phase curve, and in quadrants III and IV  it decreases, then it follows from Chaplygin's theorem that the projection of the phase curve is bounded
\begin{itemize}
\item in quadrant  I from below  by $\Lambda_{pm}$, from above  by  $\Lambda_{mp}$,
\item in quadrant  II from below  by $\Lambda_{pp}$, from above  by  $\Lambda_{mm}$,
\item in quadrant  III from below  by  $\Lambda_{mp}$, from above  by  $\Lambda_{pm}$,
\item in quadrant  IV from below  by  $\Lambda_{mm}$, from above  by  $\Lambda_{pp}$.
\end{itemize}
Thus, starting from a point, we switch the upper and lower boundaries of the trajectory, moving from one quadrant to another (see Figure 1). As the top limiter, we get a curve that unfolds around the origin. While this curve intersects  axis $ Q = 0 $ at a point $D_*$, $ 0 <D_* <\frac12 $, then, as follows from Proposition \ref{utv2.1}, the projection of the phase trajectory makes another revolution. This process can be continued until the point of intersection with axis $ Q = 0 $  in  quadrant II is greater than $ \frac12 $. If $ n $ is the number of complete revolutions, then the lifetime of the smooth solution is estimated from below as $ 2 \pi n $. Note that the fact that the curve $ \Lambda_{mm} $ intersects axis $ Q = 0 $  at $ D> \frac12 $ does not mean that the projection of the phase trajectory itself also intersects  $ Q = 0 $  at such a point. Numerical calculations show that the resulting estimate is rather rough.

We can also construct the lower bounds of the projection of the phase trajectory from the side of the origin using the curves $ \Lambda_{pm} $ and $ \Lambda_{pp} $. However, it is easy to see that they form a spiral that turns to zero, so if the conditions of Proposition \ref{utv2.2} are not met, they will not be met at the next turn. That is, no new information can be obtained.

\bigskip
{\bf Example.}
The curves $ \Lambda_{mm} $ and $ \Lambda_{mp} $, used as upper limiters for the projection of the phase trajectory, can be found explicitly by the formulas specified in \cite {RChD20}. They are rather cumbersome, so we  consider the case $ \nu = 0 $ for illustration.

  The curves $\Lambda _{mp}$ and $\Lambda _{mm}$ for $\nu=0$ have the form $C(1-D^2)=K_+ Q^2+D^2$ and $C(1-D^2)=K_- Q^2+D^2$, respectively. The constant $ C $ is first found from the initial data, and then in each next quadrant from the coordinates of the  point of intersection of the curve in the previous quadrant with the coordinate axis.

If we choose the initial data \eqref {gauss} used in the calculations in Section \ref{S4} (Subsection 4.2), that is
$$
E_0(\rho) = 0.4761 \,\rho \,\exp\left\{-0.0987654321\,
{\rho^2}\right\},\quad P_{0}(\rho) = V_{0}(\rho) = 0,
$$
we get
$n=3$ and the guaranteed lifetime of a smooth solution is $6\pi$. In this case, the smallest number of revolutions can be guaranteed for trajectories starting from points  $|\rho_0|\in (0.69, 1.2)$. 

From the results of numerics presented in Figure 2 for this case, we see that in reality, the number of oscillations before blowup is 4.

\bigskip

Note that, as $ \nu $ increases, the  curves $ \Lambda $ approach the origin, then for each fixed starting point of the trajectory, the guaranteed number of revolutions increases with $ \nu $.

\begin{center}
\begin{figure}[htb]
\includegraphics[scale=0.45]{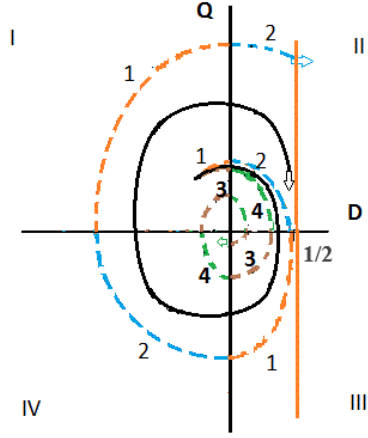}
\caption{Projection of the phase curve of  system \eqref {u1}, \eqref {char1d} onto the plane $ (D, Q) $ (solid). Bilateral limiters in each quadrant (dashed). 1 - curve $ \Lambda_{mp} $, 2 - curve $ \Lambda_{mm} $, 3 - curve $ \Lambda_{pm} $, 4 - curve $ \Lambda_{pp} $. The intersection of the upper limiter with the straight line $ D = \frac12 $ means that the continuation of the solution for the next revolution is not guaranteed.}\label{Pic1}
\end{figure}
\end{center}

\medskip

\subsection{Small deviations from the trivial state}

Previously we obtained  the following result.
\begin{theorem}\label{T2.4}\cite{RChZAMP21}
For $ \nu = 0 $, any solution to the Cauchy problem \eqref {u1}, \eqref {cd1} which is an arbitrarily small deviation from the equilibrium $ P = 0 $, $ E = 0 $, such that  $ 2 \sqrt {1 + P_0^2 (\rho)} + E_0^2 (\rho) $ is not identically equal to a constant, the derivatives  turn to infinity in a finite time.
\end{theorem}

We want to show that for an arbitrarily small collision rate, this fact is no longer true.

Note that the case
  $ 2 \sqrt {1 + P_0^2 (\rho)} + E_0^2 (\rho) =C$ for all $ \rho \in \mathbb R $ ($C_\rho=0$) corresponds to the simple wave for $ \nu= 0 $, see \eqref{SW}.

\medskip

Let us show that the following theorem holds.
\begin{theorem}\label{T2.5}
For any $ \nu> 0 $ there exists $ \epsilon> 0 $ such that the solution to the Cauchy problem \eqref {u1}, \eqref {cd1} with initial data satisfying the conditions
$ E_0^2(\rho) + P_0^2 (\rho)+ \alpha^2+\beta^2\le \epsilon^2 $ uniformly in $ \rho $ remains smooth for all $ \theta> 0 $.
\end{theorem}

\medskip

We need the following lemma.
\begin{lemma}\label{L2.6}
We denote
 $$\dfrac{\za}{\zb} = u_0, \quad
\dfrac{1-\za}{\zb} = \lambda_0, \quad \zb\ne 0.$$
Let $z(\theta)$  be a solution to the Cauchy problem
\begin{equation}\label{Hill}
\dfrac{d^2 z}{d\theta^2}-\nu \dfrac{d z}{d\theta}+K(\theta) z=0,\quad z(0)=1,\, z'(0)=-u_0.
\end{equation}
If for at least one point $ \rho_0 $ there is a moment $ \theta_*> 0 $ such that
\begin{equation}\label{y_prime}
z'(\theta_*)=\lambda_0 e^{ {\nu}\theta_*},
 \end{equation}
 then the derivatives of the solution of  \eqref {u1}, \eqref {cd1} that do not correspond to a simple wave given as \eqref{SW} blow up in a finite time.
Otherwise, the solution remains smooth for all $ \theta> 0 $.
\end{lemma}



\proof
 The change $ u = -z '/ z $ reduces the first of the equations  \eqref{supp1} to  \eqref{Hill}. This equation is homogeneous, so we set  $ z (0) = 1 $, whence   $ z '(0) = - u_0$ follows. From the second equation
 \eqref {supp1} we have $u=\lambda'/\lambda-\nu$,
and $\lambda=\frac{\lambda_0}{z} \exp \left({\nu} \theta\right)$.

 Thus,
$$
\dfrac{u}{\lambda} = \dfrac{D}{1-D} = -\,\dfrac{z'}{\lambda_0} e^{-{\nu}\theta}.
$$
If $ D $ goes to infinity at
 $\theta=\theta_*$, then
$-\,\dfrac{z'}{\lambda_0} e^{-{\nu}\theta_* }=-1$, that is, the condition \eqref{y_prime} holds.
If there is no such moment, then $ D $ and $ Q $ remain bounded.
The lemma is proved.
$\Box$


{\it Proof of Theorem \ref{T2.5}.}
 We give the proof for $ \nu \in (0,2) $, although it can be done similarly for $\nu>0$.

We start with case of general initial data and then prove the theorem for the simple wave. In order to apply the lemma, we need an explicit form $K(\theta)$. Let us make an assumption about the smallness of the initial perturbation.
 If $ P_0^2(\rho_0) + E_0^2(\rho_0)\le \epsilon^2 $,
  $ \epsilon \ll 1 $, then \eqref{char} implies that $ P $ and $ E $ remain small as long as the solution remains smooth, in particular, $ P^2 \le \epsilon^2 \ll 1 $. The function $P(\theta) $ satisfies the equation
  $${P}'' +\nu  P'+\frac{P}{\sqrt{1+P^2}}=0,$$
   therefore, up to $o(\epsilon)$
  we get $ P = \epsilon \,e^{-\frac{\nu}{2}\theta}\,\sin (\omega\theta + \theta_0) $, $\omega=\sqrt{1-\frac{\nu^2}{4}}$. Without loss of generality, we set  $ \theta_0 = 0 $, substitute $ P $ in $ K $ and expand the result in a series in $ \epsilon $. We obtain
\begin{equation}\label{K_series}
K(\theta)=1-\frac32\, e^{-\nu\theta}\,\sin^2\omega\theta \, \epsilon^2+O(\epsilon^4).
\end{equation}
 We neglect the terms of order higher than the second in \eqref{K_series}, substitute the result into \eqref {Hill} and seek a solution in the form
 $z(\theta)= z_0(\theta)+\epsilon^2 z_1(\theta)+ o(\epsilon^2)$. In the standard way we get
 \begin{eqnarray*}
   z(\theta) = e^{-\frac{\nu}{2}\theta}\left(\frac{\nu/2-u_0}{\omega}\sin \omega\theta+\cos \omega\theta\right)+\\
   \left[ e^{ \frac{\nu}{2}\theta} A_1\sin (\omega\theta+\theta_1)+
     e^{-\frac{3\nu}{2}\theta}(A_2 \sin(\omega\theta+\theta_2)+A_3 \sin (3\omega\theta+\theta_3)\right]\,\epsilon^2+o(\epsilon^2),
  \end{eqnarray*}
 where $A_i=A_i(\nu, u_0),$  $\theta_i=\theta_i(\nu, u_0),$ $i=1, 2, 3$.

 Let us fix $\nu>0$. For $ \theta_* $ to exist (see Lemma \ref{L2.6}), there must be a positive root of the equation
 \begin{equation}\label{theta_s}
   \lambda_0 e^{\nu\theta}= z'(\theta)=  e^{-\frac{\nu}{2}\theta} \Theta_1 + \epsilon^2 [ e^{\frac{\nu}{2}\theta} \Theta_2 +   e^{-\frac{3\nu}{2}\theta} \Theta_3]+o(\epsilon^2),
 \end{equation}
  where $\Theta_i=\Theta_i(\nu, u_0, t) $, $i=1, 2, 3$,  are bounded functions. The absolute value of the left hand side of \eqref{theta_s} grows as $ \theta \to \infty $ faster than the absolute value of the right hand side, so a possible root can be only for $ \theta $, belonging to the first period of $\Theta_1$. The amplitude of $\Theta_1$ is $A_{\Theta_1}=\left(\frac{\left(\frac{\nu}{2}u_0-1\right)^2}{\omega^2}+u_0^2\right)^{1/2}$. Thus, if condition
  $A_{\Theta_1}<|\lambda_0|$ holds,
   then as for
$\theta>0$ the amplitude of the $ \epsilon $-independent expression on the right hand side decreases, then by decreasing $ \epsilon $,
we can always ensure that the graph of the growing exponential function on the left side does not intersect with the graph of the function on the right side. This ensures that there is no root of  equation \eqref{theta_s}.
The condition $A_{\Theta_1}<|\lambda_0|$ implies
\begin{equation*}\label{DQ}
  \omega^2 (2\alpha^2-1)+(\beta-\frac{\nu}{2}\alpha)^2<0,
\end{equation*}
therefore for sufficiently small $\alpha^2+\beta^2$ this inequality holds. Thus, the theorem is proved for the case of general initial data.

For the case of a simple wave \eqref{char1d} and \eqref{SW} result in
\begin{equation}\label{SWQ}
\dfrac{dQ}{d\theta}=-Q \left(K(P) Q +\nu-\frac{P}{\sqrt{1+P^2}(E(P)+\nu P)} \right).
  \end{equation}
We can see that $K(P)=1+O(P)$, $\frac{P}{\sqrt{1+P^2}(E(P)+\nu P)} = O(P)$, therefore for any fixed $\nu>0$ there exist
sufficiently small $P^2_0(\rho_0)+Q^2_0(\rho_0)$ such that the solution to \eqref{SWQ} tends to zero as $\theta\to\infty$.
Now Theorem \ref{T2.5} is proved for all cases.
$\Box$

\section
{ Numerical illustrations}\label{S4}

The approach to the analysis of characteristics used to obtain analytical results can be transformed into a high-precision approximate method for calculating relativistic oscillations of cold plasma with allowance for collisions. The specificity of the method lies in the identification of the displacement function of particles (electrons) when using Lagrangian variables with a function that characterizes the electric field when using Eulerian variables.

In fact, the method presented below is a method of characteristics; however, it is convenient to present its construction in terms of the Lagrangian description of the medium, i.e. using the concepts of particles and their trajectories.

Note that the simulated medium is represented by charged particles located in the background field formed by stationary ions. Therefore, for each particle there is an "equilibrium" position when the background field is canceled and which is convenient to take as
values of its Lagrangian coordinate. In this case, the displacement function relative to the equilibrium position is responsible for the formation of the electric field:
\begin{equation}\label{ptraek}
\rho(\rho^L,\theta) = \rho^L + R(\rho^L,\theta),
\end{equation}
where $ \rho^L $ is the "equilibrium" position of the particle when it does not contribute to the formation of the electric field,
$ R (\rho^L, \theta) $ is its displacement function generating an electric field at the point of the trajectory $ \rho (\rho^L, \theta) $.
According to \cite{daw59}, in the flat one-dimensional case, there is a simple relationship between the field and the displacement
\begin{equation}
\label{peq3}
R(\rho^L,\theta) = E(\rho,\theta) \equiv E(\rho^L + R(\rho^L,\theta),\theta),
\end{equation}
which makes the functions of the electric field and displacement on the particle trajectory indistinguishable.
Note that this approach additionally makes it possible not to solve simultaneously two identical differential equations with different initial conditions (to determine the Eulerian trajectory of a particle and the electric field along it).
In this case, the formal transition from the Euler coordinates $ (\rho, \theta) $ to the Lagrangian coordinates $ (\rho^L, \tau) $ is carried out by the usual transformation
$$
\tau \equiv \theta, \quad \rho^L = \rho - \int\limits_{\tau_0}^{\tau} d\tau'V(\rho^L,\tau')\,,
$$
but instead of $ \tau $ the old notation $\theta$ is used .

\subsection
{ Numerical algorithm}

To find a numerical solution on the line $ \rho \in (- \infty, \infty) $, we define an arbitrary grid at the initial moment of time $\theta=0$:
$$
          \rho_{1}(0) < \rho_{2}(0) < \dots < \rho_M(0),
$$
consisting of $ M $ nodes. To each node $ \rho_k (\zt = 0), \; 1 \le k \le M, $ place a particle,
marked with the Lagrangian coordinate $ \rho^L_k $.
Moreover, for each particle from \eqref {u1} and \eqref {peq3}, there follow equations describing the dynamics of particles in Lagrangian variables:
\begin{equation}
\label{peq4}
\begin{array}{c}
\dfrac{d \,P(\rho^L_k,\theta) }{d \,\theta} = - R(\rho^L_k,\theta) - \nu P(\rho^L_k,\theta), \quad
\dfrac{d \,R(\rho^L_k,\theta)}{d \,\theta} = V(\rho^L_k,\theta), \\
\quad V(\rho^L_k,\theta) = \dfrac{P(\rho^L_k,\theta)}{\sqrt{1+P^2(\rho^L_k,\theta)}}\,,\quad k =1, 2, \dots, M.
\end{array}
\end{equation}

Let us use the equalities \eqref {ptraek} and \eqref {peq3} to obtain the missing initial conditions for systems of the form \eqref {peq4}. At the node $ \rho_k $ for $ \zt = 0 $, $ E_0 (\rho_k) $ is defined, i.e. $ \rho_k = \rho^L_k + E_0 (\rho_k) $ holds, whence the initial conditions follow for the particle with the number $ k $:
 \begin{equation}
\label{peq5}
P(\rho^L_k,\theta=0)= P_0(\rho_k), \quad R(\rho^L_k,\theta=0)= E_0(\rho_k), \quad k = 1,2, \dots, M,
\end{equation}
as well as the equilibrium value of the Lagrangian coordinate $ \rho^L_k $ itself. The obtained relations allow, instead of problem \eqref {u1}, \eqref {cd1}, written in Eulerian variables,
solve numerically the problem \eqref {peq4}, \eqref {peq5} formulated in Lagrangian variables.

The solution of  problem \eqref {peq4}, \eqref {peq5} determined in this way does not allow finding the spatial derivatives of the required functions $ P $ and $ E $, which makes it impossible to determine the electron density function $ N $ in accordance with \eqref {3gl4}. To avoid this drawback, it is enough to reformulate the problem \eqref {du1}, \eqref {dcd1} from Eulerian variables to Lagrangian variables, using the reasoning presented above. Formal transformations give the equations
\begin{equation}
\label{peq6}
\begin{array}{l}
\dfrac{d \,Q(\rho^L_k,\theta) }{d \,\theta} = - D(\rho^L_k,\theta) - \dfrac{W^2(\rho^L_k,\theta)}{\left(1+P(\rho^L_k,\theta)^2\right)^{3/2}}
- \nu Q(\rho^L_k,\theta), \vspace{1.0 ex} \\
\dfrac{d \,D(\rho^L_k,\theta)}{d \,\theta} = (1 -  D(\rho^L_k,\theta))\, \dfrac{Q(\rho^L_k,\theta)}{\left(1+P(\rho^L_k,\theta)^2\right)^{3/2}},
\end{array}
\end{equation}
and initial conditions
 \begin{equation}
\label{peq7}
Q(\rho^L_k,\theta=0)= Q_0(\rho_k), \quad D(\rho^L_k,\theta=0)= D_0(\rho_k),
\end{equation}
corresponding to individual particles with numbers $k = 1,2, \dots, M.$

Now, from the solution of \eqref{peq6}, \eqref {peq7} and the relation \eqref {ptraek} for each particle
 \begin{equation}
\label{peq8}
\rho_k(\zt) = \rho^L_k + R(\rho^L_k,\theta), \quad k =1, 2, \dots, M,
\end{equation}
it is possible at a Eulerian point of space $ (\rho_k (\zt), \theta) $ to determine the value of the electron density
$$
N(\rho_k, \theta) = 1 - D(\rho^L_k,\theta).
$$
Thus, the proposed numerical algorithm consists in finding solutions of  equations \eqref {peq4} with  conditions \eqref {peq5} at the nodes of the Eulerian grid \eqref {peq8}, as well as  equations \eqref {peq6} with  conditions \eqref {peq7} for each of the particles numbered $ k = 1, 2, \dots, M. $
Note that the search for an approximate solution presupposes the existence, uniqueness, and smoothness of the exact solution of the problem due to Propositions \ref{utv2.1} - \ref{utv2.3}, and, as noted in Section \ref{S1}, general theorems on the existence and uniqueness of a locally smooth solution to the Cauchy problem for systems of hyperbolic equations \cite {Daf16}, pp.221-225.
If the solution has sufficient smoothness in the variable $ \zt $, it seems very convenient to use
the classical Runge - Kutta method of the fourth order of accuracy \cite {KMN89}, otherwise (of less smoothness) one should use schemes of a lower order of accuracy up to the Euler method.
It is necessary to clarify that in this situation the accuracy of the obtained approximation is determined exclusively by the smoothness of the solution.

Note also that the stability of the time integration of the equations for the momentum and the electric field is completely determined by the inequality \eqref{first_int}.
There is no analogous property for equations describing the dynamics of their derivatives; therefore, under initial conditions that do not lead to the intersection of Lagrangian trajectories in the non-relativistic case, in the relativistic case these trajectories can intersect.
This situation leads to a discontinuity in the function $ E (\rho, \zt) $, and, as a consequence, the singularity
electron density functions in accordance with \eqref{3gl4}. Therefore, in the process of calculations, it is necessary to constantly monitor the preservation of the starting order of particles, i.e. conditions $ \rho_{k} (\zt^n) <\rho_{k + 1} (\zt^n) $ for all $ k = 1, 2, \dots, M-1. $
Violation of this condition for at least one value of $ k $ means a blow-up (impossibility of further application of the model), denoted by the term "breaking".
The above means that observing or investigating a solution with specific initial data only makes sense until the moment of blow-up.

For this reason, further we will keep in mind that the numerical method is applied on a limited time interval $ \zt \in [0, \zt_ {br}] $, while the desired solution to the problem \eqref {3gl4} - \eqref {cd1} exists , is unique and bounded together with the necessary derivatives.
In other words, it is further assumed that the order of Lagrangian particles does not change on the segment $ [0, \zt_{br}] $, that is, the inequality
$$
h_k(\zt) = \rho_{k+1}(\zt) - \rho_{k}(\zt) > 0 \quad \forall k =1,2, \dots, M-1,
$$
holds
together with the sufficient condition for a local in the time  existence (see Proposition \ref{utv2.1})
\begin{equation*}\label{critrel}
 \left(\dfrac{\partial P(\rho,\theta)}{\partial \rho}\right)^2 + 2\, \dfrac{\partial E(\rho,\theta)}{\partial \rho} - 1 < 0.
\end{equation*}

\medskip

Summarizing the above description of the numerical algorithm, we emphasize that for solving the original problem \eqref{3gl4} - \eqref {dcd1} written in Eulerian variables, it is convenient to go over to Lagrangian variables and use approximate methods of integration over time to calculate the required functions at the points $ \rho_k (\zt^n), \quad k = 1, 2, \dots, M, $ belonging to the trajectories of particles. This is quite enough for studying most of the problems. However, it is quite possible that situations arise when it is required to determine the solution at the given Eulerian points $ (\rho, \theta) $, which do not have to belong to the calculated trajectories of the particles.

In this case, at the moment of time $ \zt^n $ it is necessary to determine first the interval $ [\rho_k (\zt^n), \rho_{k + 1} (\zt^n)] $,
to which the given value $ \rho $ belongs, and then use the interpolation procedure to find the approximate value.

Taking into account that at the nodes of the Euler grid $ \rho_k (\zt ^ n), \quad k = 1, 2, \dots, M, $ not only the functions $ P (\rho, \zt) $ and $ E ( \rho, \zt) $ are used, but also their spatial derivatives, it seems very convenient to use the Hermitian cubic interpolation for this purpose. The derivation of the necessary formulas and error estimates is given in \cite {Schultz73} pp.24-39, the practical details of use, including the necessary programs, are well described in~\cite {KMN89}, pp.110-114.


\subsection
{ Calculation results}
As the initial conditions, we choose the functions
\begin{equation}\label{gauss}
E_0(\rho) = \left(\dfrac{a_*}{\rho_*}\right)^2\rho \exp\left\{-2
\dfrac{\rho^2}{\rho_*^2}\right\},\quad P_{0}(\rho) = V_{0}(\rho) = 0.
\end{equation}
According to \eqref {gauss} and \eqref {3gl4}, initially the density minimum is at the origin, and the density deviations from the equilibrium  equal to one decay exponentially.
This choice is the most natural from the point of view of physics, since it simulates the effect on a rarefied plasma of a short powerful laser pulse when it is focused into a line (this can be achieved using a cylindrical lens), see details in \cite{Shep13}.

As the working area of localization of oscillations, we choose the segment $ [- d, d] $ at $ d = 4.5 \, \rho_* $.
In this case, the amplitudes of particle oscillations at the boundaries will coincide in order of magnitude with the machine accuracy, due to
$\exp\{-2d^2/\rho_*^2\} \approx 2.58 \cdot 10^{-18}$, that is, the near-boundary characteristics will coincide with the straight lines $\rho \approx \pm d$ with a high degree of accuracy.
It is convenient to choose a uniform initial grid  $\rho_k(\zt=0) = kh - d,\; k =0,1, \dots, M,\, h=2d/M.$

As an illustration of the algorithm, we present the calculations of the blow-up effect that can be observed
after several periods of oscillations. We fix the parameters $ a_* = 3.105, $ $ \rho_* = 4.5 $ in order to satisfy the condition of a local in time existence of a solution for  system \eqref {u1} from Proposition \ref{utv2.1},
and also to preserve the continuity with the results of numerical and asymptotic analysis from~\cite {FrCh}.
The calculations for the extended system \eqref {u1} - \eqref {dcd1} were carried out by the classical Runge - Kutta method of the 4th order of accuracy at $ \tau = h = 0.01 $ until the intersection of the trajectories of Lagrangian particles, which in the Euler formulation corresponds to an infinite electron density. In order to control the accuracy, calculations were additionally carried out with grid parameters two times smaller than the main (working) ones.

The results of calculations for the electron density as a function of time at various parameters $ \nu $, which characterizes the damping of oscillations due to collisions of electrons and ions, are shown in Figures 2 - 4. Figure 2 shows the time dependence of the density maximum in a collisionless plasma at $ \nu = 0 $, i.e. in the absence of damping. It follows from Figure 1 that for the given calculation parameters in the fourth oscillation period, a density maximum is formed outside the oscillation axis, which already in the next period at $ \zt^{(0)}_{wb} \approx 29.5 $ has an infinitely large value.

\begin{center}
\begin{figure}[htb]
\includegraphics[scale=1.0]{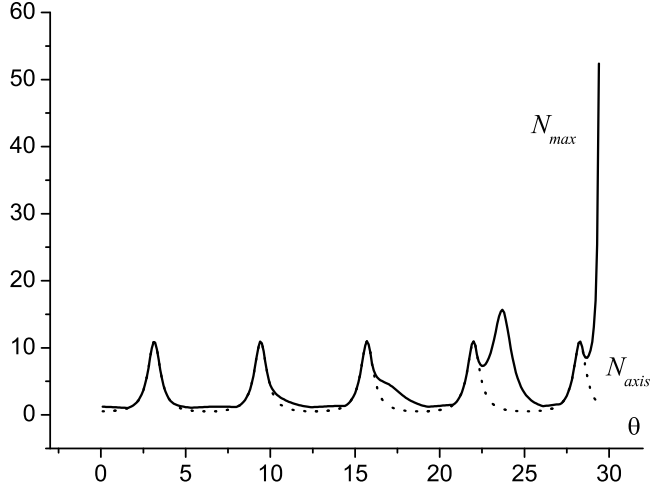}
\caption{Time dependence of the electron density in nonlinear oscillations of a collisionless plasma ($ \nu \zt^{(0)}_{wb} = 0 $). The maximum density in the entire computational domain (solid) and the density at the origin (dashed).
}\label{Pic1}
\end{figure}
\end{center}

\begin{center}
\begin{figure}[htb]
\includegraphics[scale=1.0]{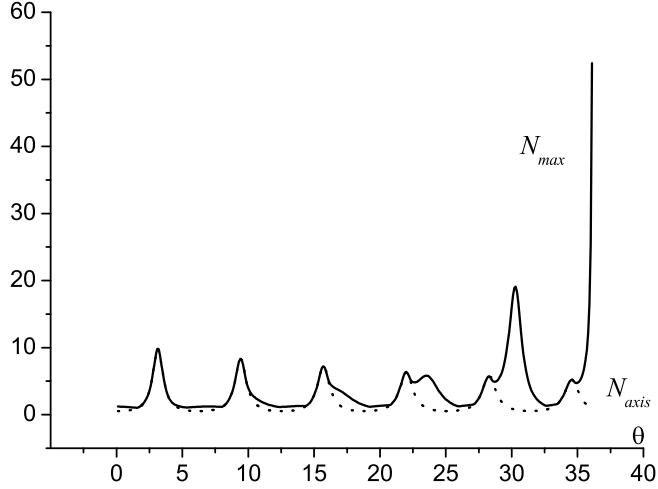}
\caption{Time dependence of the electron density in nonlinear plasma oscillations taking into account collisions ($ \nu \zt^{(0)}_{wb} = 0.2 $). The maximum density in the entire computational domain (solid), the density at the origin (dashed).
}\label{Pic2}
\end{figure}
\end{center}

\begin{center}
\begin{figure}[htb]
\includegraphics[scale=1.0]{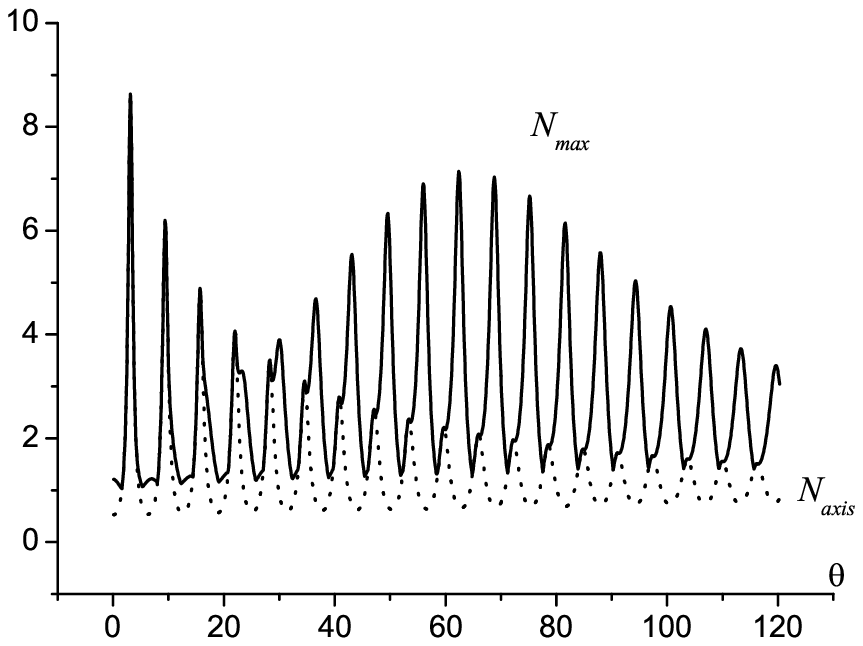}
\caption{Time dependence of the electron density in nonlinear plasma oscillations with frequent collisions ($ \nu \zt^{(0)}_{wb} = 0.5$). The maximum density in the entire computational domain (solid), the density at the origin (dashed).
}\label{Pic3}
\end{figure}
\end{center}

When the collisions of electrons are taken into account, the time for the appearance of the density singularity increases. For example, for $ \nu \zt^{(0)}_{wb} = 0.2 $, the off-axis maximum (see Figure 2) has an infinite value only in the third period after its formation, and not in the second period as it was in collisionless case. Calculations show that the density singularity for given initial parameters arises only in the case of relatively rare collisions, when the inequality
$ \nu \zt^{(0)}_{wb} \le 0.422 $. Note that the dependence of the breaking time on the parameter $ \nu \zt^{(0)}_{wb} $ with allowance for collisions was established by asymptotic methods in~\cite {FrCh}. When the equality $ \nu \zt^{(0)}_{wb} = 0.422 $ holds, the time for the density singularity to appear is $ \zt_{wb} \approx 75.22 $, and it is about 2.5 times the breaking time in a collisionless plasma. When the condition $ \nu \zt^{(0)}_{wb}> 0.422 $ is satisfied, the density singularity does not arise anymore. This scenario of the evolution of the electron density maximum is shown in Figure 8 for $ \nu \zt^{(0)}_{wb} = 0.5 $. In this case, the off-axis maximum, after its formation, first increases, but then it decreases due to the strong damping of oscillations.
Figure 5 shows the spatial distributions of the momentum $ P $ and the electric field $ E $ of a collisionless plasma at the moment of blow-up $ \theta \approx 29.5 $, when the absolute  maximum of density outside the origin became unbounded, and the trajectories of a pair of neighboring particles intersect. Note that, due to the structure of  equations \eqref {u1}, their solution will remain odd functions of coordinates if the initial data have this property. The initial data \eqref{gauss} is exactly that, therefore, graphs of functions are shown only on the positive semiaxis.

It is easy to see that in the vicinity of the density maximum, the velocity component has a discontinuity in the derivative (a weak discontinuity), but not in the function itself, while a strong discontinuity is formed in the electric field function. It is these qualitative characteristics of $ P $ and $ E $ that ensure the breaking (blow-up) of oscillations. Let us emphasize that blow-up has the character of a "gradient catastrophe", i.e. the functions $ P $ and $ E $ themselves remain bounded.
\begin{center}
\begin{figure}[htb]
\includegraphics[scale=1.0]{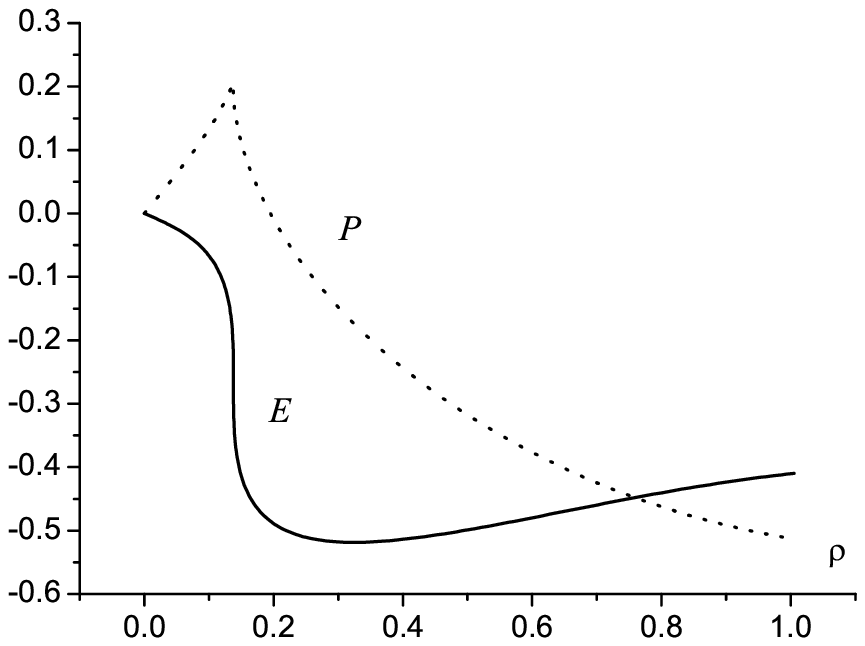}
\caption{Spatial distribution of momentum and electric field at the moment of overturning in a relativistic calculation of a collisionless plasma ($ \nu \zt^{(0)}_{wb} = 0$).
}\label{Pic1}
\end{figure}
\end{center}
Note that the above results of calculations by the particle method were completely reproduced by additional calculations according to the splitting scheme in Euler variables and according to the time hopping scheme ("leap - frog") in Lagrange variables (see details in~\cite{Ch_book}).

Let us present estimates for the breaking time of nonlinear oscillations for some typical plasma parameters ~\cite{FrCh}. If plasma oscillations are driven by  electric field \eqref {gauss} with parameters $ a_* = 3.105, $ $ \rho_* = 4.5 $, then the blowup time in a collisionless plasma is $ \zt^{(0)}_{wb} \approx 29.5 $, as follows from the calculations. Hence, in accordance with the numerical result for the threshold value of the dimensionless collision frequency $ \nu \zt^{(0)}_{wb} \le 0.422 $, it follows that the breaking effect takes place when the following inequality is satisfied $ \nu \le 1.43 \cdot 10^{- 2} $. In a fully ionized plasma, the dimensionless frequency of electron-ion collisions is determined by the formula~\cite{SR12}
 \begin{equation}\label{fornu}
\nu = Z \dfrac{\sqrt{8}}{3}\eta^{3/2} \ln \Lambda,
\end{equation}
where $ Z $ is the ion charge number, $ \ln \Lambda $ is the Coulomb logarithm, and the parameter $ \eta $ is equal to the ratio of the interaction energy of electrons $ e^2 N_{0e}^{1/3} $ to their kinetic energy $T_e$
 \begin{equation}\label{foreta}
\eta =  \dfrac{e^2 N_{0e}^{1/3}}{T_e}.
\end{equation}
Let in a rarefied, fully ionized plasma with an ion charge $ Z = 5 $, whose electrons have a density $ N_{0e} = 10^{18} {\rm cm}^{- 3} $ and a temperature $ T_e = 50 $ $eV$, a laser pulse propagates with a wavelength $ \lambda = 1.24 $ $\rm \mu m$ (frequency $ \omega_0 \approx 1.5 \cdot 10^{15} {\rm s}^{- 1} $), duration $ \tau \approx 36 $ fs and dimensionless electric field amplitude $ a_0 \approx 2.5 $. If a laser pulse is focused by a cylindrical lens into a line with a transverse size $ L_x \approx 24 $ $\rm \mu m$, then in the wake wave behind the pulse, mainly plane one-dimensional oscillations of electrons with parameters $ a_* = 3.1, $ $ \rho_* = 4.5 $ are excited, which are close to ones used in the calculations above. Note that when a laser pulse with a moderately relativistic intensity $ a_0$ (basically between 1 and 3) propagates in a rarefied plasma, the condition for optimal excitation of plasma waves ($ \tau_* = 2 $) is approximately conserved, and the amplitude of plasma oscillations is related to the laser field by the same relation $ a_*^2 \approx 1.52 \, a_0^2 $, as in the non-relativistic limit. For the reduced plasma parameters, from  formulas \eqref{fornu}, \eqref{foreta} we find that the dimensionless collision frequency $ \nu \approx 0.5 \cdot 10^{- 2} $ is less than the threshold value $ 1.43 \cdot 10^{- 2 } $. Therefore, in this case, the breaking effect takes place, and the breaking time is approximately equal to $ \zt_{wb} \approx 33 $, since the parameter $ \nu \zt^{(0)}_{wb} \approx 0.15 $. If we consider the propagation of a laser pulse with the above parameters in a plasma with the same density, but with a temperature of $ T_e = 20 $ eV, then calculations using  formulas \eqref{fornu}, \eqref{foreta} give the dimensionless collision frequency $ \nu \approx 1.8 \cdot 10^{- 2} $ (parameter $ \nu \zt^{(0)}_{wb} \approx 0.52 $) that exceeds the threshold. Therefore, in this case, strong damping suppresses the breaking of plasma oscillations in the laser pulse wakefield. 

\vspace{1.5em}
\section
{ Conclusion}
In the present work, the influence of collisions of electrons on the breaking (blow-up) of plane nonlinear plasma oscillations is investigated analytically and numerically. If there are no collisions of electrons in the plasma, then the breaking of plasma oscillations occurs due to the formation of a maximum electron density outside the oscillation axis, which increases with time and goes to infinity after several periods of oscillations. It is shown that, when electron collisions are taken into account, the breaking is also associated with an increase in this off-axis density maximum, but in a collisional plasma its growth occurs more slowly, and the breaking time increases with an increase in the collision frequency. It has been established analytically and numerically that for any fixed initial data there is a certain threshold value for the frequency of electron collisions, upon reaching which the density singularity does not arise. It is shown that, if the collision frequency above the threshold value, the density maximum outside the oscillation axis after its formation increases only for a certain time, but then it decreases due to strong damping of oscillations, i.e. the solution is stabilized to the trivial state.

\section*{Acknowledgment}
Partially supported by the Moscow Center for Fundamental and Applied Mathematics.

\end{document}